\documentclass[twocolumn]{svjour3}          
\smartqed  
\usepackage{graphicx}
\begin{document}

\title{Exotic Spin-Orbital Physics in Hybrid Oxides
}


\author{Wojciech Brzezicki \and Mario Cuoco \and Andrzej M. Ole\'s
}


\institute{Wojciech Brzezicki
          \and
           Mario Cuoco \at
             CNR-SPIN, IT-84084 Fisciano (SA), Italy, and
             Dipartimento di Fisica \textquotedblleft{}E. R. Caianiello\textquotedblright{},
             Universit\'a degli Studi di Salerno, IT-84084 Fisciano (SA), Italy
           \and
           Andrzej M. Ole\'s \at
             a.m.oles@fkf.mpg.de \\
             Marian Smoluchowski Institute of Physics, Jagiellonian University,
             prof. S. \L{}ojasiewicza 11, PL-30348 Krak\'ow, Poland;
             Max Planck Institute for Solid State Research,\\
             Heisenbergstrasse 1, D-70569 Stuttgart, Germany
}

\date{Received: date / Accepted: date}

\maketitle

\begin{abstract}
We compare the effective spin-orbital super\-exchange triggered by
magnetic $3d$ impurities with $d^3$ and $d^2$ configurations and
either no orbital degree of freedom (orbital dilution) or hole
replacing a doublon (charge dilution) in a $4d^4$ Mott insulator
with $S=1$ spins. Impurities causing orbital dilution act either
as spin defects decoupled from the surrounding ions, or generate
orbital polarons along $d^3$-$d^4$ hybrid bonds. The exchange
on these bonds determines which orbital is occupied by a doublon
on the host site.
In case of charge dilution by $3d^2$ impurities additional 
$\propto T_i^+T_j^+$ terms arise which enhance orbital fluctuations.
We show that such terms may radically change orbital pattern at
relatively low doping by $x=1/8$ hole defects. Our findings provide
new perspective for future theoretical and experimental studies of
doped transition metal oxides.

\end{abstract}

\section{ Spin-orbital physics }

Spin-orbital physics was initiated by Kugel and Khomskii \cite{Kug82}
who realized that orbital operators contribute to superexchange in
Mott insulators in a similar way as spins, so both degrees of freedom
contribute jointly to spin-orbital superexchange \cite{Ole05}. While
the spin part has SU(2) symmetry, the orbital part has remarkable
orbital fluctuations \cite{Kha05} and a much lower cubic symmetry in
the perovskite systems where it is intrinsically frustrated 
\cite{Ole12}. This frustration is frequently released by spin order 
which coexists with orbital order following the Goodenough-Kanamori 
rules \cite{Gee96}, and phases arise with alternating orbital (AO) 
order along ferromagnetic (FM) bonds coexisting with ferro-orbital 
(FO) order along antiferromagnetic (AF) bonds.
Well known examples are spin-orbital orders in LaMnO$_3$ \cite{Fei99},
LaVO$_3$ \cite{Kha04}, and Ca$_2$RuO$_4$ \cite{Cuo06} where spin-orbit
coupling plays a role \cite{Fiona} --- the latter two examples (V, Ru)
involve $t_{2g}$ orbitals. Indeed, a rather unique example of a
spin-orbital system are perovskite vanadates, where a challenging
competition between two types of spin-orbital order was observed
\cite{Fuj10}. But a different scenario is also possible
--- frustrated spin-orbital interactions may lead to the collapse of
long-range order, as for instance in LiNiO$_2$ \cite{Rei05}.
Another possibility is a spin-orbital liquid emerging from
frustration \cite{Nor08,Karlo}.

Doping of Mott insulators leads to several remarkable phenomena.
Recently short-range charge density wave called stripe phase was
reported in doped cuprates \cite{Cam15}. When coupled to spins it
arises as self-organization of charge and spin degrees of freedom
\cite{Tra96}. It has been suggested that the critical charge, orbital,
and spin fluctuations near the quantum critical point provide the
pairing interaction \cite{Bia00} and the spectral properties of stripe
phases may be seen as a signature of their stability \cite{Fle01}.
In manganites hole doping generates orbital polarons that emerge in an
AF system by double-exchange mechanism \cite{Kil99}, responsible for a
change from AF insulator to FM metal with $e_g$ orbital liquid
\cite{Fei05}.

As in doped cuprates, the holes doped in $e_g$ orbitals are mobile in
LaMnO$_3$, or in Kugel-Khomskii system \cite{Tan09}, as
well as in $t_{2g}$ stripe phases with orbital polarons \cite{Wro10}.
More complex phenomena are found in doped vanadates where a $t_{2g}$
hole may hop only in two cubic directions \cite{Dag08}, and robust 
$C$-type AF ($C$-AF) coexists with $G$-type AO ($G$-AO) order
\cite{Fuj08}. Doping generates finite spectral weight within the
Mott-Hubbard gap already at low $x\simeq 0.02$ Ca doping in 
Y$_{1-x}$Ca$_x$VO$_3$ \cite{Ave13}, while at higher doping the 
system remains insulating and develops a kinetic gap \cite{Ave15}.

The purpose of this paper is to consider the consequences of
\textit{immobile holes} in a ruthenate due to transition metal ions
with a lower valence which are substituted for Ru ($d^4$) ions.
As an experimental motivation we mention that:
($i$) dilute Cr doping for Ru reduces the temperature of the
orthorhombic distortion and induces FM order in
Ca$_2$Ru$_{1-x}$Cr$_x$O$_4$ (with $0<x<0.13$) \cite{Qi10}, and
($ii$) Mn-substituted single crystals of Sr$_3$Ru$_{2-x}$Mn$_x$O$_7$
reveal an unusual $E$-type AF structure at $x=0.16$ \cite{Mes12}.
Therefore we consider:
($i$) \textit{orbital dilution} by $3d^3$ (Mn$^{4+}$, Cr$^{3+}$)
impurities in (Ca,Sr)$_2$RuO$_4$ \cite{Brz15}, and
($ii$) \textit{charge dilution} due to $3d^2$ doping which generates
hole-doublon interactions. These findings motivate the theoretical
search for the consequences of both orbital and charge dilution. We
show that spin-orbital order may even be globally changed and dictated
by defects in case of orbital dilution with $x=1/4$ \cite{Brz16}. Note
that in contrast to orbital dilution in cuprates \cite{Tan09} where
holes remove both spin and orbital degree of freedom, the present
doping 
($i$) removes the orbital doublon but increases spin from $S=1$
to $S=3/2$, while in 
($ii$) spins $S=1$ stay unchanged but a doublon is
replaced by a hole in an orbital (holon).

\section{ Orbital dilution }

The spin-orbital model for Ca$_2$RuO$_4$ Mott insulator
is equivalent by an electron-hole transformation to that introduced
some 15 years ago for vanadates \cite{Kha01}. Thus we consider an
interplay between $S=1$ spins and the $t_{2g}$ orbital doublet
$T=1/2$ for a doublon active along a given cubic axis. For instance,
the doublon involves $\{yz,zx\}$ active orbitals along the $c$ axis,
while $xy$ orbitals are filled by one electron each. The model is
isomorphic to the vanadate $d^2$ model \cite{Kha04}, with doublons
transforming into empty orbitals (occupied by two holes). We label
$t_{2g}$ orbitals by index $\gamma$ when a given orbital is inactive
along a cubic axis $\gamma\in\{a,b,c\}$:
\begin{equation}
\left|a\right\rangle\equiv\left|yz\right\rangle, \hskip .5cm
\left|b\right\rangle\equiv\left|xz\right\rangle, \hskip .5cm
\left|c\right\rangle\equiv\left|xy\right\rangle.
\label{eq:or_defs}
\end{equation}

\begin{figure*}[t!]
\begin{center}
\includegraphics[width=16.1cm]{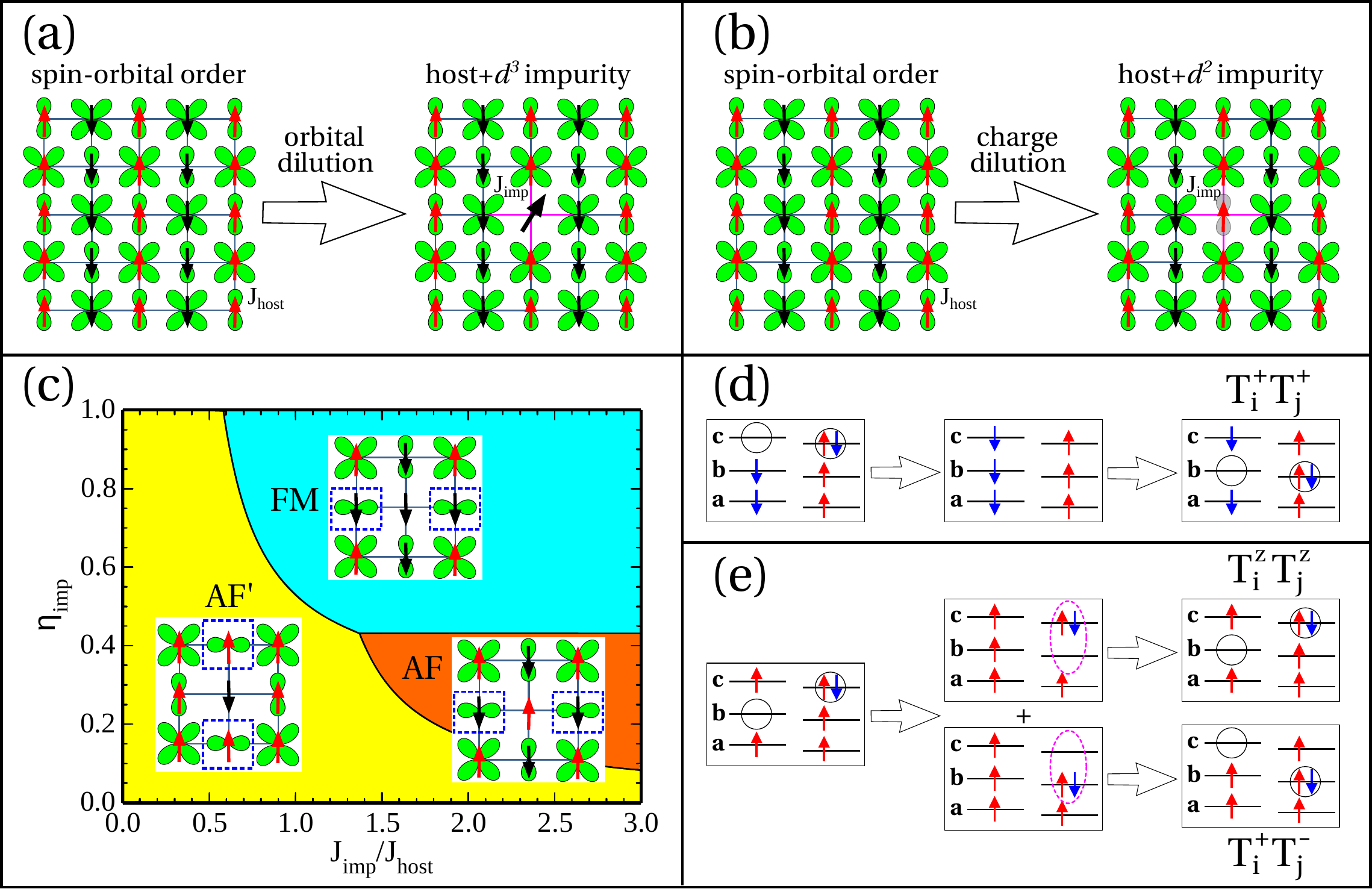}
\end{center}
\caption{
Top --- Doping by transition metal ions in a spin-orbital system with
$C$-AF and $G$-AO $\{a,c\}$ order found in Mott insulators
with $d^2$ (vanadates) or $d^4$ (ruthenates) ionic configuration:
(a) orbital dilution with $a$ orbital removed by the $d^3$ impurity
with $S=3/2$ spin, and
(b)  charge dilution with gray orbital indicating an
orbital hole at the $d^2$ impurity with $S=1$ spin.
Host $S=1$ spins interacting by $J_{\rm host}$ are shown by red/black
arrows and doublons in $t_{2g}$ orbitals ($a$ and $c$) are shown by
green symbols. Here doping occurs at $a$ doublon site and spins are
coupled by $J_{\rm imp}$ along hybrid (red) bonds.
Bottom left --- 
(c) phase diagram for a single $d^3$ impurity replacing a doublon in 
$c$ orbital in the $C$-AF host \cite{Brz15}; dashed boxes indicate
change in the orbital order around the impurity by the orbital flip
$a\rightarrow b$.
Bottom right --- 
Orbital fluctuations promoted on $d^2$-$d^4$ hybrid bonds with:
(d) AF, and
(e) FM spin correlations.
In the latter case (e) the terms $\propto J_2$ couple doublons at two
orbitals in excited states (doublon and hole in ovals), and
one obtains orbital flips $\propto T_i^-T_j^+$
together with Ising terms $\propto T_i^zT_j^z$,
while  double excitations $\propto T_i^+T_j^+$ occur on AF bonds (d) 
even in the absence of Hund's exchange (at $J_2=0$) and are amplified 
by finite $\{J_1,J_2\}$. Splittings between 
degenerate $t_{2g}$ orbitals serve only as guides for the eye.
}
\label{fig:bond}
\end{figure*}

We consider a two-dimensional (2D) RuO$_2$ $ab$ plane in Ca$_2$RuO$_4$
(SrRuO$_3$), with Ru ions connected via $2p_{\pi}$ oxygen orbitals.
For this plane $|a\rangle$ ($|b\rangle$) orbitals are active along the
$b$ ($a$) axis, while $|c\rangle$ orbitals are active along both $a$
and $b$ planar axes. The superexchange for the host bonds 
$\langle ij\rangle$ along the $\gamma\in\{a,b\}$ axis \cite{Cuo06},
\begin{equation}
{\cal H}_{d^4-d^4}=J_{\rm host}\sum_{\langle ij\rangle\parallel\gamma}
\left\{J_{ij}^{(\gamma)}(\vec{S}_{i}\!\cdot\!\vec{S}_{j}+1)J_{ij}
+K_{ij}^{(\gamma)}\right\},
\label{eq:Hhost}
\end{equation}
is $\propto J_{\rm host}$ and stabilizes $C$-AF phase for the realistic
parameters of Ca$_2$RuO$_4$, see Fig. \ref{fig:bond}(a). The above
generic form \cite{Ole12} follows from charge excitations,
$4d^4_i4d^4_j\Rightarrow 4d^5_i4d^3_j$ ---
the interactions depend on the intraorbital Coulomb $U_2$ element and
Hund's exchange $J_2$ in the host. The orbital operators,
$J_{ij}^{(\gamma)}$ and $K_{ij}^{(\gamma)}$, are defined by active
$t_{2g}$ orbital along the axis $\gamma$ \cite{Cuo06}. For a bond
$\langle ij\rangle\parallel\gamma$ it may be rewritten in the
orbital-only form by taking average over spin degrees of freedom, i.e.,
\begin{equation}
{\cal H}_{d^4-d^4}^{\langle ij\rangle\parallel\gamma}=
\left\{A_{\gamma}^{}T_i^zT_j^z+
\frac12 C_{\gamma}^{}\left(T_i^+T_j^-+T_i^-T_j^+\right)\right\}^{(\gamma)},
\label{eq:H44}
\end{equation}
with orbital operators $(T_i^{\alpha})^{(\gamma)}$ defined by the axis
$\gamma$ and with coefficients $A_{\gamma}$ and $C_{\gamma}$ depending
on spin average $\langle\vec{S}_i\!\cdot\!\vec{S}_j\rangle$ on a bond
$\langle ij\rangle$.

For the $d^3$-$d^4$ hybrid bonds charge
$3d^3_i4d^4_j\Rightarrow 3d^4_i4d^3_j$ excitations with the lowest 
energy do not generate
an extra doublon but simply move it from $4d$ to $3d$ ion. We use the
convention that $i=1$ stands for the impurity site and $i=2$ for its
host neighboring ion. The energy involved in the charge excitation is,
\begin{equation}
\Delta=I_{e}+3(U_{1}-U_{2})-4(J_{1}-J_{2})\,.
\label{Delta}
\end{equation}
where $U_i$ and $J_i$ are the respective Coulomb and Hund's elements.
In addition it depends on the ionic energy $I_e$, i.e., the mismatch
between the energy levels of the two atoms. For Mn or Cr
impurities in ruthenates $\Delta>0$ and plays a role of charge-transfer
energy. The energy $\Delta$ (\ref{Delta}) defines two parameters which
characterize the interactions along the hybrid $d^3$-$d^4$ bonds
\cite{Brz15}:
\begin{equation}
J_{\rm imp}=\frac{t^{2}}{4\Delta}, \hskip 1.2cm
\label{eq:etai}
\eta_{\rm imp} =\frac{J_{1}}{\Delta}.
\label{eq:imp} 
\end{equation}

The spin-orbital interaction on hybrid bond has the generic form
similar to Eq. (\ref{eq:Hhost}), but the orbital operators are now
defined only by the host site, and the spin operators are for $S=1$
on host and $S=3/2$ on impurity ion. A more transparent form of the
impurity-host bond couples the impurity spin $\vec{S}_i$ with the
neighboring host spin $\vec{S}_j$ --- it can be written as follows
\cite{Brz15},
\begin{equation}
{\cal H}_{d^3-d^4}^{\langle ij\rangle\parallel\gamma}\simeq
\left\{J_{S}(D_j^{(\gamma)})(\vec{S}_i\!\cdot\!\vec{S}_j)
+ E_D^{\gamma} D_j^{(\gamma)}\right\}.
\label{eq:H123}
\end{equation}
Here the spin exchange couplings $J_S(D_j^{(\gamma)})$ depend on
doublon projection operator $D_j^{(\gamma)}$ at host site $j$, and the
doublon energy $E_D^{\gamma}$ depends on Hund's exchange
$\eta_{\rm imp}$ Eq. (\ref{eq:imp}). It can be shown \cite{Brz15} that
the latter is the dominant energy scale, so for a single $d^3$-$d^4$
bond the doublon avoids the inactive ($\gamma$) orbital and spins
couple with $J_{S}(D_j^{(\gamma)}=0)$ which can be either AF if
$\eta_{\rm imp}<0.43$, or FM if $\eta_{\rm imp}>0.43$. The sign change
at $\eta_{\rm imp}^c\simeq 0.43$ marks a quantum phase transition from
AF to FM spin correlations, see Fig. \ref{fig:bond}(c). Similar to the 
2D Kugel-Khomskii model \cite{Brz12}, this sign change leads here to 
rather exotic phases with nearly frustrated spins for several doping 
levels $x\in[1/9,1/5]$ \cite{Brz15}. A single $d^3$ impurity at site 
$i$ modifies the spin-orbital order at its nearest neighbors (NNs)
$j\in{\cal N}(i)$, while second NNs are little affected
and thus they typically follow the $C$-AF/$G$-AO order in the host.
In particular, 
the impurity spin reorients within the orbital polaron at
$\eta_{\rm imp}^c$ which marks a transition from AF to FM regime.

Both for a single impurity and at low $x\le 1/8$ doping, the $d^3$-$d^4$
bonds influence strongly spin-orbital order \cite{Brz15}. For a higher
periodic doping $x=1/4$ when half of the superexchange bonds are
$d^3$-$d^4$ hybrid bonds, the overall spin-orbital order is dictated by
them \cite{Brz16}. One finds that only every second undoped vertical
line $\parallel b$ is FM, as in the $C$-AF host phase, and host spins
are inverted on any other vertical line and the doublon flips from 
orbital $a$ to $b$. Such a modification of the orbital order stabilizes
the FM interactions for $c$-$a$ doublon pairs on the horizontal bonds
by double-exchange, in analogy to a doped $t_{2g}$ system \cite{Wro10}.
In the phase diagram at $x=1/4$ doping \cite{Brz16} one finds a broad
region of parameters where impurity spins are also frustrated.
Frustration is released by quantum fluctuations which stabilize 
impurity spin orientation opposite to that expected in the host. At 
sufficiently large $\eta_{\rm imp}$ (\ref{eq:imp}), FM spin order 
takes over and $G$-AO order is then the same as in the undoped host.
Thus, the most interesting novel spin-orbital phases are indeed found 
at the crossover from AF to FM interactions along hybrid bonds.

\begin{figure}[t!]
\begin{center}
\includegraphics[width=\columnwidth]{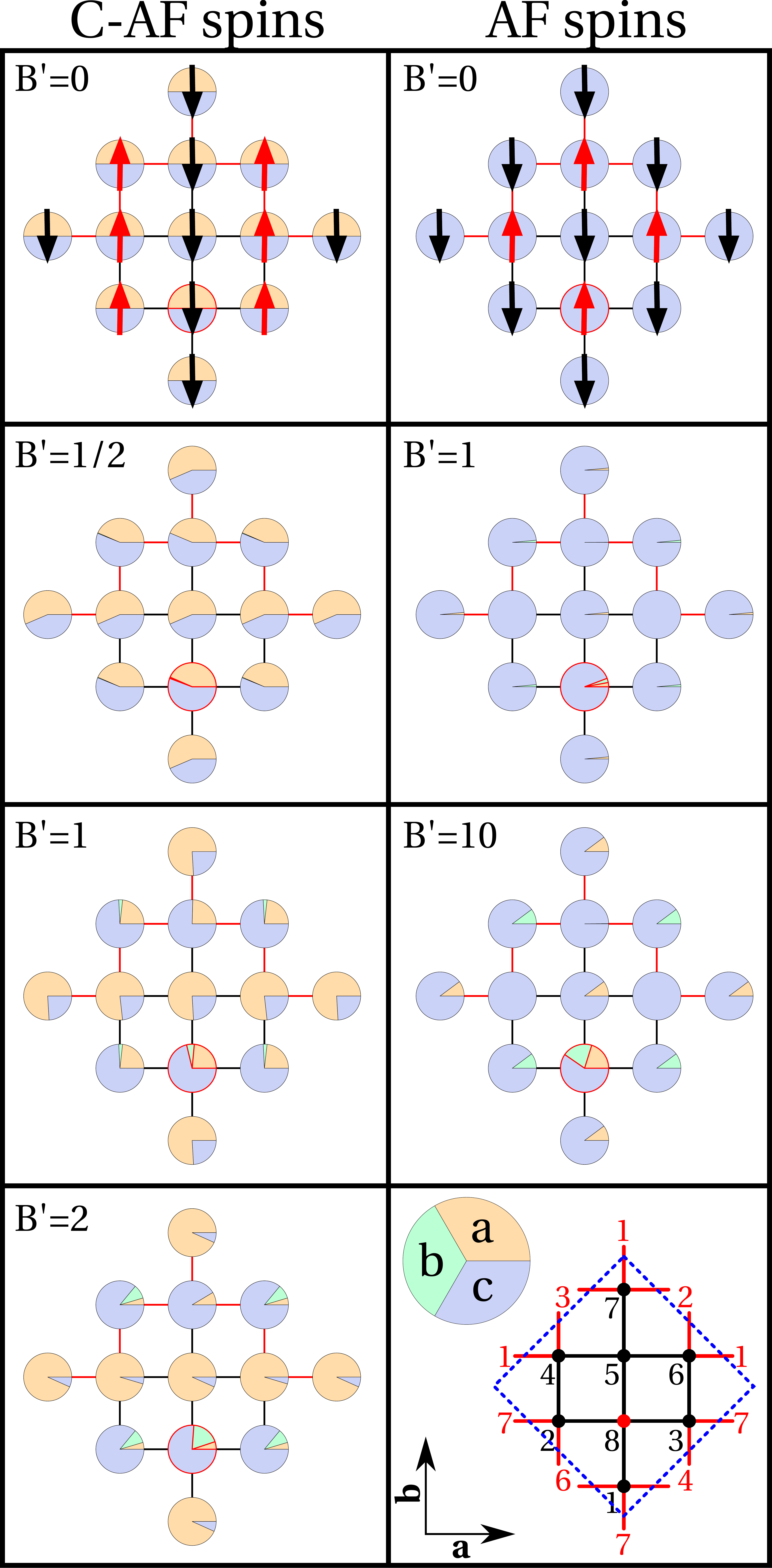}
\end{center}
\caption{
Orbital patterns obtained at $x=1/8$ charge dilution (with periodic
cluster of $N=8$ sites) in two AF phases in an $ab$ plane:
left  --- $C$-AF phase with FM vertical ($\parallel b$) lines, and
right --- N\'eel AF phase (shown by arrows for $B^{'}=0$).
Orbital fluctuations on the bonds around the impurity at site $i=8$
are $\propto B'$ (\ref{eq:H24}) and increase from top to bottom. 
Bottom right --- 
color convention for an orbital occupied by a doublon at host sites 
and a hole at impurity site (red circles), as well as the cluster of 
$N=8$ sites (dotted square) with the periodic boundary conditions on 
outer (red) bonds.}
\label{fig:of}
\end{figure}

\section{ Charge dilution }

Doping a spin-orbital $t_{2g}^4$ system by $t_{2g}^2$ ions is even
more challenging. In this case the impurity itself has another orbital
degree of freedom, a hole (holon) in place of a doublon. A particular
feature of $d^2$ ions is that they promote orbital fluctuations in
case when $c$ orbital ($xy$) is filled by one electron at all sites,
and the other two $\{a,b\}$ ($\{yz,zx\}$) fluctuate along the $c$ axis,
see Eqs. (\ref{eq:or_defs}).
This supports FM spin order even in the absence of Hund's exchange
\cite{Kha01}. It is thus challenging to consider such ions within a
ruthenate where similar orbital fluctuations are also expected.

The generic form of spin-orbital superexchange (\ref{eq:Hhost})
applies as well to the hybrid $d^2$-$d^4$ bonds where charge
$3d^2_i4d^4_j\Rightarrow 3d^3_i4d^3_j$ excitations may generate an 
extra doublon in intermediate states at a $3d$ ion for AF bonds (not 
shown), or instead they create a high-spin state along FM bonds, see 
Fig. \ref{fig:bond}(d). As a result, the spin-orbital interactions are 
here quite complex and include several terms, so we  do not reproduce 
them here. They are controlled again by the charge-transfer energy and 
the same effective parameters as for orbital dilution, see Eqs. 
(\ref{eq:imp}). Again, the spin exchange depends on whether the doublon 
at the host ion is in the active or inactive orbital on the considered 
$d^2$-$d^4$ bond. It also depends on the configuration at the impurity 
site, i.e., whether the holon is in an active or inactive orbital.

The derivation of spin-orbital superexchange for the hybrid $d^2$-$d^4$
bonds demonstrates a remarkable difference to the orbital dilution
on $d^3$-$d^4$ bonds discussed above. Consider first the AF spin bond
shown in Fig. \ref{fig:bond}(d). Charge excitations,
$3d^2_i4d^4_j\Rightarrow 3d^3_i4d^3_j$, create, \textit{inter alia},
two high-spin configurations which generate novel orbital quantum 
fluctuations, $\propto T_i^+T_j^+$, which go beyond those found in the 
host \cite{Kha01}. In addition, a doublon may be created at the $3d$ 
impurity ion and such configurations have to be combined with others 
to obtain eigenstates at both sites at finite $\{J_1,J_2\}$. 
As a result, the fluctuations $\propto T_i^+T_j^+$ are amplified. 
They are accompanied by Ising orbital exchange terms and the usual 
fluctuations $\propto T_i^+T_j^-$ activated by finite Hund's exchange 
(not shown). For a FM spin configuration the doublon gets a partner 
empty orbital at the host site and again, orbital quantum fluctuations 
$\propto T_i^-T_j^+$ arise at finite $J_2$, see Fig. \ref{fig:bond}(e).

Altogether, after averaging spin correlations, the purely orbital 
superexchange is,
\begin{eqnarray}
{\cal H}^{\langle ij\rangle\parallel\gamma}_{d^2-d^4}&=&
\left\{A_{\gamma}^{'}T_i^zT_j^z+
\frac12 C_{\gamma}^{'}\left(T_i^+T_j^-+T_i^-T_j^+\right)\right.
\nonumber\\
&+&\left.\frac12 B_{\gamma}^{'}\left(T_i^+T_j^++T_i^-T_j^-\right)
\right\}^{(\gamma)},
\label{eq:H24}
\end{eqnarray}
with orbital operators $(T_i^{\alpha})^{(\gamma)}$ defined by the bond
direction and the coefficients $A_{\gamma}^{'}$, $B_{\gamma}^{'}$, and
$C_{\gamma}^{'}$. As these parameters for the present hybrid bond are
similar to those in the host (\ref{eq:H44}), we use below,
\begin{equation}
A_{\gamma}^{'}=A_{\gamma},  \quad
B_{\gamma}^{'}\equiv B^{'}, \quad
C_{\gamma}^{'}=C_{\gamma},
\label{ABC}
\end{equation}
i.e., we characterize the orbital superexchange on the hybrid bonds by 
the parameter $B^{'}$ which stands for double excitations in Fig. 
\ref{fig:bond}(d). Due to such terms the orbital quantum number 
${\cal T}^z$ is not conserved.

A more complete analysis of charge dilution and the resulting phase
diagrams will be presented elsewhere. Here we focus on the consequences
of orbital fluctuations $\propto B^{'}$, both in $C$-AF and N\'eel AF
spin configurations in the $ab$ plane, see Fig. \ref{fig:of}. Quantum
fluctuations in the spin subsystem are small for $S=1$ spins and we
neglect them. For realistic Hund's exchange in the host and at the
holon ion one expects $C$-AF order. Its orbital pattern at $B^{'}=0$
has equal orbital densities in $a$ and $c$ orbitals, see left column
in Fig. \ref{fig:of}. Indeed, this configuration guarantees maximal
doublon fluctuations along the $b$ axis where the order is FM, as in 
the vanadium perovskites with holon fluctuations involving $a$ and 
$b$ orbitals along the $c$ axis \cite{Kha01}. Increasing orbital 
fluctuations gradually destroy this optimal state and favor AO order 
along vertical FM columns, with enhanced $a$ or $c$ holon density at 
every other site. For small $B^{'}=1/2$ the $T_i^+T_j^+$ term is active 
only along $b$-bonds and consequently there is no admixture of $b$ 
orbitals in the ground state. At larger $B^{'}=1$ we see the appearance 
of the $b$ orbital polarization at the impurity site and two neighboring
host sites along $a$ axis which goes together with a severe change of 
the global orbital order. At $B^{'}=2$ a new state appears with almost 
complete separation into horizontal chains of $a$ ($c$) orbitals
and a small admixture of $b$ orbital density. Holon site has a
similar orbital density distribution, with somewhat enhanced $b$
character. We suggest that the orbital fluctuations of alternating
$a$ and $c$ orbitals along the vertical bonds replace here double
exchange to stabilize FM spin order, while such fluctuations for the
same orbitals along the horizontal bonds support AF spin order.

Actually, FO order in the $ab$ plane supports AF N\'eel state following
the Goodenough-Kanamori rules \cite{Gee96}, see the right column in Fig.
\ref{fig:of}. This state is more robust and almost unchanged by double
excitations $\propto T_i^+T_j^+$ with a realistic $B^{'}=1$, and only
at a high value of $B^{'}=10$ one finds increased $a$ or $b$ orbital
densities following the checkerboard pattern around the impurity. This
effect seems to be purely local because already the second neighbors of
the impurity in the host are unaffected even for such a large value of 
$B^{'}$. Unlike in $C$-AF spin order, we see that the defects in the
orbital configurations are strongly localized or screened by the host.
This screening has an interesting geometrical mechanism: if we look
around the impurity we see that the up and down neighbors get an
admixture of the $a$ orbitals from the $\propto T_i^+T_j^+$ terms 
induced by the impurity. These $a$-orbital defects cannot however 
delocalize in the $a$ direction (along two such bonds). The defects can 
only move up, to the site number $7$ in the cluster (see bottom right 
of Fig. \ref{fig:of}). On the other hand, the horizontal neighbors of 
the impurity are in analogical situation with $b$-orbital defects which 
for the same reason can only go to the same site number $7$. Finally, 
at this site one expects a destructive interference of the $a$-orbital 
defects going from the top and bottom bonds and the $b$-orbital defects 
going from the left and right ones as it exhibits only the initial 
$c$-orbital polarization. It is thus possible that for a different 
distribution of impurities we would observe a much weaker localization 
of the defects.

\section{ Discussion and summary }

Orbital (charge) dilution plays a role in several Mott insulators with
spin-orbital order doped by transition metal ions with either $d^3$ or
$d^2$ electron configuration. We have shown that the orbital order
around such impurities changes in general, so even in the dilute limit
one may expect observable effects such as islands of reversed spins or
doublon fluctuations. We argue that the general trends reported here
are generic and a meaningful insights into phase diagrams of systems
with orbital dilution may be gained by investigating classical phase
diagrams as quantum fluctuations are small \cite{Brz15}. Double
exchange leads to local or global changes of spin-orbital order,
similar to orbital polarons \cite{Kil99} or dimensional crossover with
a change from $G$-AF to $C$-AF order in electron doped manganites 
\cite{Ole11}. Doping may also generate novel spin-orbital-charge 
modulated patterns reported recently for $t_{2g}$ systems \cite{Prl15}.

Summarizing, this study highlights the role played by orbital or charge
dilution by transition metal ions in cubic spin-orbital systems and
provides new insights necessary for theoretical understanding of
(Mn,Cr)-doped layered ruthenates and related systems. Previous
studies \cite{Brz15} have shown that $d^3$ impurities change radically
spin-orbital order, and we expect an even stronger impact of $d^2$
impurities on spin-orbital order in the $d^4$ host --- the changes
may be strongly centered around the impurity sites, as shown in the
FO case, or completely smeared out by the host's fluctuations,
as in the $C$-AF case. The theoretical studies of this
doping are under way. It is challenging to investigate the
consequences of enhanced orbital fluctuations in experimental systems.

\begin{acknowledgements}
Open access funding provided by Max Planck Society.
W.B. acknowledges support by the European Union's Horizon 2020
research and innovation programme under the Marie Sklodowska-Curie
grant agreement No. 655515.
We acknowledge support by Narodowe Centrum Nauki
(NCN, National Science Center), Project 2012/04/A/ST3/00331.
\end{acknowledgements}


\begin{thebibliography}{99}

\bibitem{Kug82} Kugel, K.I., Khomskii, D.I.:
                   The Jahn-Teller effect and magnetism:
                   Transition metal compounds,
                   Usp. Fiz. Nauk \textbf{136}, 621 (1982)
                  [Sov. Phys. Usp. \textbf{25}, 231 (1982)]

\bibitem{Ole05} Ole\'s, A.M., Khaliullin, G., Horsch, P., Feiner, L.F.:
                   Fingerprints of spin-orbital physics in cubic Mott insulators:
                   Magnetic exchange interactions and optical spectral weights.
                   Phys. Rev. B \textbf{72}, 214431 (2005)

\bibitem{Kha05} Khaliullin, G.:
                   Orbital order and fluctuations in Mott insulators.
                   Prog. Theor. Phys. Suppl. \textbf{160}, 155 (2005)

\bibitem{Ole12} Ole\'s, A.M.:
                   Fingerprints of spin-orbital entanglement
                   in transition metal oxides.
                   J. Phys.: Condens. Matter \textbf{24}, 313201 (2012)

\bibitem{Gee96} Geertsma, W, Khomskii, D.I.:
                   Influence of side groups on 90 degrees superexchange:
                   A modification of the Goodenough-Kanamori-Anderson rules.
                   Phys. Rev. B \textbf{54}, 3011 (1996)

\bibitem{Fei99} Feiner, L.F., Ole\'s, A.M.:
                   Electronic origin of magnetic and orbital ordering
                   in insulating LaMnO$_3$.
                   Phys. Rev. B \textbf{59}, 3295 (1999)

\bibitem{Kha04} Khaliullin, G., Horsch, P., Ole\'s, A.M.:
                   Theory of optical spectral weights
                   in Mott insulators with orbital degrees of freedom.
                   Phys. Rev. B \textbf{70}, 195103 (2004)

\bibitem{Cuo06} Cuoco, M., Forte, F., Noce, C.:
                   Interplay of Coulomb interactions and $c$-axis
                   octahedra distortions in single-layer ruthenates.
                   Phys. Rev. B \textbf{74}, 195124 (2006)

\bibitem{Fiona} Cuoco, M., Forte, F., Noce, C.:
                   Probing spin-orbital-lattice correlations
                   in $4d^4$ systems.
                   Phys. Rev. B \textbf{73}, 094428 (2006)

\bibitem{Fuj10} Fujioka, J., Yasue, T., Miyasaka, S., Yamasaki, Y.,
                   Arima, T., Sagayama, H., Inami, T., Ishii, K., Tokura, Y.:
                   Critical competition between two distinct
                   orbital-spin ordered states in perovskite vanadates.
                   Phys. Rev. B \textbf{82}, 144425 (2010)

\bibitem{Rei05} Reitsma, A., Feiner, L.F., Ole\'s, A.M.:
                   Orbital and spin physics in LiNiO$_2$ and NaNiO$_2$.
                   New J. Phys. \textbf{7}, 121 (2005)

\bibitem{Nor08} Normand, B., Ole\'s, A.M.:
                   Frustration and entanglement in the $t_{2g}$
                   spin-orbital model on a triangular lattice:
                   Valence-bond and generalized liquid states.
                   Phys. Rev. B \textbf{78}, 094427 (2008)

\bibitem{Karlo} Corboz, P., Lajk\'o, M., La\"uchli, A.M., Penc, K., Mila, F.:
                   Spin-orbital quantum liquid on the honeycomb lattice,
                   Phys. Rev. X \textbf{2}, 041013 (2012)

\bibitem{Cam15} Campi, G., Innocenti, D., Bianconi, A.:
                   CDW and similarity of the Mott insulator-to-metal
                   transition in cuprates with the gas-to-liquid-liquid
                   transition in supercooled water.
                   J. Supercond. Nov. Magn. \textbf{28}, 1355 (2015)

\bibitem{Tra96} Tranquada, J.M., Axe, J.D., Ichikawa, N., Nakamura, Y.,
                   Uchida, S., Nachumi B.:
                   Neutron-scattering study of stripe-phase order of holes
                   and spins in La$_{1.48}$Nd$_{0.4}$Sr$_{0.12}$CuO$_4$.
                   Phys. Rev. B \textbf{54}, 7489 (1996)

\bibitem{Bia00} Bianconi, A., Bianconi, G., Caprara, S., Di Castro, D.,
                   Oyanagi H., Saini, N.L.:
                   The stripe critical point for cuprates.
                   J. Phys.: Condens. Matter \textbf{12}, 10655 (2000)

\bibitem{Fle01} Fleck, M., Lichtenstein, A.I., Ole\'s, A.M.:
                   Spectral properties and pseudogap in the stripe phases
                   of cuprate superconductors.
                   Phys. Rev. B \textbf{64},  134528 (2001)

\bibitem{Kil99} Kilian, E., Khaliullin, G.:
                   Orbital polarons in the metal-insulator transition
                   of manganites.
                   Phys. Rev. B \textbf{60}, 13458 (1999)

\bibitem{Fei05} Feiner, L.F., Ole\'s, A.M.:
                   Orbital liquid in ferromagnetic manganites:
                   The orbital Hubbard model for $e_g$ electrons.
                   Phys. Rev. B \textbf{71}, 144422 (2005)

\bibitem{Tan09} Tanaka, T., Ishihara, S.:
                   Dilution effect in correlated electron systems
                   with orbital degeneracy.
                   Phys. Rev. B \textbf{70}, 035109 (2009)

\bibitem{Wro10} Wr\'obel, P., Ole\'s, A.M.:
                   Ferro-orbitally ordered stripes in systems
                   with alternating orbital order.
                   Phys. Rev. Lett. \textbf{104}, 206401 (2010)

\bibitem{Dag08} Daghofer, M., Wohlfeld, K., Ole\'s, A.M., Arrigoni, E.,
                   Horsch, P.:
                   Absence of hole confinement in transition-metal oxides
                   with orbital degeneracy.
                   Phys. Rev. Lett. \textbf{100}, 066403 (2008)

\bibitem{Fuj08} Fujioka, J., Miyasaka, S., Tokura, Y.:
                   Doping variation of anisotropic charge and
                   orbital dynamics in Y$_{1-x}$Ca$_x$VO$_3$:
                   Comparison with La$_{1-x}$Sr$_x$VO$_3$.
                   Phys. Rev. B \textbf{77}, 144402 (2008)

\bibitem{Ave13} Avella, A., Horsch, P., Ole\'s, A.M.:
                   Defect states and excitations in a Mott insulator with
                   orbital degrees of freedom: Mott-Hubbard gap versus
                   optical and transport gaps in doped systems.
                   Phys. Rev. B \textbf{87}, 045132 (2013)

\bibitem{Ave15} Avella, A., Ole\'s, A.M., Horsch, P.:
                   Defects, disorder, and strong electron correlations
                   in orbital degenerate, doped Mott insulators.
                   Phys. Rev. Lett. \textbf{115}, 206403 (2015)

\bibitem{Qi10}  Qi, T.F., Korneta, O.B., Parkin, S., De Long, L.E.,
                   Schlottmann, P., Cao, G.:
                   Negative volume thermal expansion via orbital and magnetic
                   orders in Ca$_2$Ru$_{1-x}$Cr$_x$O$_4$ ($0<x<0.13$).
                   Phys. Rev. Lett. \textbf{105}, 177203 (2010)

\bibitem{Mes12} Mesa, D., Ye, F., Chi, S., Fernandez-Baca, J.A.,
                   Tian. W., Hu, B., Jin, R., Plummer, E.W., Zhang, J.:
                   Single-bilayer $E$-type antiferromagnetism in Mn-substituted
                   Sr$_3$Ru$_2$O$_7$: Neutron scattering study.
                   Phys. Rev. B \textbf{85}, 180410(R) (2012)

\bibitem{Brz15} Brzezicki, W., Ole\'s, A.M., Cuoco, M.:
                   Spin-orbital order modified by orbital dilution
                   in transition-metal oxides: From spin defects
                   to frustrated spins polarizing host orbitals.
                   Phys. Rev. X \textbf{5}, 011037 (2015)

\bibitem{Brz16} Brzezicki, W., Cuoco, M., Ole\'s, A.M.:
                   Novel spin-orbital phases induced by orbital dilution.
                   J. Supercond. Nov. Magn. \textbf{29}, 563 (2016)

\bibitem{Kha01} Khaliullin, G., Horsch, P., Ole\'s, A.M.:
                   Spin order due to orbital fluctuations: Cubic vanadates.
                   Phys. Rev. Lett. \textbf{86}, 3879 (2001)

\bibitem{Brz12} Brzezicki, W., Dziarmaga, J., Ole\'s, A.M.:
                   Noncollinear magnetic order stabilized
                   by entangled spin-orbital fluctuations.
                   Phys. Rev. Lett. \textbf{109}, 237201 (2012)

\bibitem{Ole11} Ole\'s, A.M., Khaliullin, G.:
                   Dimensional crossover and the magnetic transition
                   in electron doped manganites.
                   Phys. Rev. B \textbf{84}, 214414 (2011)

\bibitem{Prl15} Brzezicki, W., Noce, C., Romano, A., Cuoco, M.:
                   Zigzag and checkerboard magnetic patterns in
                   orbitally directional double-exchange systems.
                   Phys. Rev. Lett. \textbf{114}, 247002 (2015)


\end{thebibliography}
\end{document}